\documentclass[aps,twocolumn,amsmath,amssymb]{revtex4}
\usepackage{graphicx}
\usepackage[usenames, dvipsnames]{color}
\usepackage{url}
\usepackage[ pdftex, plainpages = false, pdfpagelabels, 
         pdfpagelayout = useoutlines,
         bookmarks,
         bookmarksopen = true,
         bookmarksnumbered = true,
         breaklinks = true,
         linktocpage=all,
         pagebackref=false,
         colorlinks = true,
         linkcolor = BrickRed,
         urlcolor = blue,
         citecolor = BrickRed,
         anchorcolor = green,
         hyperindex = true,
         hyperfigures
         ]{hyperref}
\usepackage{color}

\begin{document}

\title{Power and Leadership: A Complex Systems Science Approach \\ Part I---Representation and Dynamics}
\author{Yaneer Bar-Yam}
\affiliation{New England Complex Systems Institute, 277 Broadway, Cambridge MA 02139}
\date{October 21, 2018}

\begin{abstract}
Historical social narratives are dominated by the actions of powerful individuals as well as competitions for power including warfare, revolutions, and political change. Advancing our understanding of the origins, types and competitive strength of different kinds of power may yield a scaffolding for understanding historical processes and mechanisms for winning or avoiding conflicts. Michael Mann introduced a framework distinguishing four types of power: political, military, economic, and ideological. We show this framework can be justified based upon motivations of individuals to transfer decision making authority to leaders: Desire to be a member of a collective, avoiding harm due to threat, gaining benefit due to payment, acquiring a value system. Constructing models of societies based upon these types of power enables us to distinguish between social systems and describe their dynamics. 
Dynamical processes include (a) competition between power systems, (b) competition between powerful individuals within a power system of a society, and (c) the dynamics of values within a powerful individual. A historical trend in kinds of power systems is the progressive separation of types of power to distinct groups of individuals. In ancient empires all forms of power were concentrated in a single individual, e.g. Caesar during the Pax Romana period. In an idealized modern democratic state, the four types of power are concentrated in distinct sets of individuals. The progressive separation of the types of power suggests that in some contexts this confers a ``fitness'' advantage in an evolutionary process similar to the selection of biological organisms. However, individual countries may not separate power completely. The influence of wealth in politics and regulatory capture is a signature of the dominance of economic leaders, e.g. the US. Important roles of political leaders in economics and corruption are a signature of the dominance of political leaders, e.g. China. Ideological leaders dominate in theocracies, e.g. Iran. Military leaders dominate in dictatorships or countries where military leaders play a role in the selection of leaders, e.g. Egypt. 
\end{abstract}

\maketitle

\section{Introduction to power and group activity}
Social power is the ability to control or influence the behaviors of others. History is often described through the interaction between individuals exerting power over one another. Power relations among a set of people determine whether they are considered as a group or political entity. Sociologists have been concerned with the underlying nature of power, how it arises, how it affects those that have it and those that don't, its roles in regulating relationships between people in groups ranging from households to global civilization, and categories of power \cite{hobbesleviathan,john1959bases,galbraith1983anatomy,mann1984autonomous,french2001bases,keltner2003power,fiske2007social}.  

Michael Mann \cite{michael1986sources,mann1993sources,mann2012sources3,michael2012sources4} developed a framework in which there are four types of power: ideological, economic, military and political. In his approach these are loosely connected societal networks of control. Ideological power results from the ability to control meaning in society because people have a need to define meaning, norms, and ritual practice. Examples include religions as well as secular ideologies such as Marxism. Economic power derives from control over the satisfaction of subsistence needs through the extraction, transformation, distribution and consumption of the objects of nature. Political power derives from the usefulness of centralized, institutionalized, territorialized regulation of many aspects of social behavior. Military power derives from the necessity of organized physical defense and its usefulness for aggression, and more specifically concentrated lethal violence.  Mann provides an extensive historical discussion of examples of how these forms of power occur and relate to each other over history. 

Here we formalize the framework developed by Mann by showing that it is a ``universal" framework for representing power due to the inherent structure of individual motivation. Mann did not find social theory discussions of motivations starting from basic human drives (e.g. sexual fulfillment, affection, health, physical exercise and creativity, intellectual creativity and meaning, wealth, prestige, power, etc) useful. Our approach more directly considers the mechanisms and dynamics of power in group activities. The objective is to understand how multiple individuals together perform activities in which an individual, or small group of individuals, determines what is done. Motivation at the individual level is directly understood from the perspective of the individual who yields rather than the individual who exercises power. The individual who yields power must have a motivation for the cognitive act of yielding control over actions, expecting a better condition from doing so than would otherwise occur. We understand these forms of power as motivated respectively by: the perceived benefit from membership in a group---political control; the avoidance of direct harm---military power; the receipt of direct benefit---economic power; and change in values self-consistently motivating that change---ideological power. On the one hand, this discussion relates to the traditional political science of elites that are able to influence or determine the behavior of groups. On the other hand, this is an extension of a complex systems characterization of the mechanisms by which collective behaviors arise \cite{dcs,multiscalevariety}. 

\section{From individual to collective behavior}

Social systems are built out of coordination of individuals into collective action. One of the important mechanisms for the creation of such behaviors is through control by one individual over the actions of others. This is a traditional mechanism by which collective action arises, distinct from self-organized behaviors such as those of swarms and flocks which can arise without an individual exercising control over the collective. The latter has been the subject of much of complex systems study of collective action due to its limited treatment in prior work. Here we extend complex systems concepts to the more traditional topic of control and power. 

Examples of behaviors that involve multiple individuals performing the same tasks in unison include marching soldiers in a Roman legion, or coordinated actions in a factory. A central question is identifying what model of an individual is needed in order to identify the mechanisms of collective action. Our objective is not to identify the actions that are performed but rather the mechanisms of control over actions.

Considering an individual in isolation may not serve as a basis of description of collective behaviors in many contexts. In this case, however, there is a meaningful primitive concept: an individual in isolation performs actions for self-benefit. Among these actions are food gathering or shelter construction. How such individually directed actions are transformed in a social context to become collective behaviors, that may or may not benefit the individual, other specific individuals, or society as a whole, is the question we are addressing. 

Mechanisms of control and the actions that they give rise to may or may not be evaluated as positive or negative within a value system (i.e. power may be used to constructively coordinate or to coerce and exploit). This is not the central question we are discussing, though some identification of the way types of power are interpreted in value systems is appropriate once we have identified the mechanisms.   

In considering the structure of power systems it is useful to identify different types of power. Distinguishing a few types cannot be a complete specification of their diversity, but can provide a basis for discussion of the dynamics of power and the evolution of social systems including, but not limited to, governance structures. Our analysis will show that the four types of power identified by Mann \cite{michael1986sources}---political, military, economic and ideological---are immediately justified in this context. 

The development of a theoretical framework enables inferences about various properties of power and its dynamics. Given different forms of power, their interactions over time are an important part of the dynamics of power historically. The consequences of dominance of one individual or the dominance of one type of power over others can be seen historically. 

An example of such a dynamic is embodied in the expression `power corrupts,' which interestingly suggests that power can be used for either collective or individual benefits but generally will result in the individual ones predominating over time. In considering that there are multiple forms of power, we can reasonably hypothesize and observe historically that when one type of power dominates other types over time more negative consequences for the public arise as well as outsized benefits for those in power. By comparison, 
a balance over different kinds of power, and dynamic power shifting, yields benefits for the public. We can understand the reasons for this observation from an analysis of dominance and dynamics.

This paper is organized as follows: In Section III we frame social power as the transfer of authority over decision making. Section IV provides an abstract formal introduction to the four types of power. Section V describes political power, Section VI military and economic power, and Section VII ideological power. Section VIII discusses the role of institutions in power. Section IX and Section X describe the representation of power graphically. Section XI introduces a notation for power systems. Section XII describes the use of historical data to discuss power systems. Section XIII describes a selection of historical examples. Section XIV describes the dominance of power types. Section XV describes the dynamics of power. Section XVI provides concluding summary. 

\section{From physical to social power: force acting over distance to transfer authority}

It is interesting to begin by considering the distinction between power as it is understood in physics and in social systems. Power in physics is a force exercised over a distance. We can understand the power to change the movement of an object in this fashion. The greater the power, the larger the change in momentum (a change in speed or direction of motion of an object) that can be achieved, where other forces are not present. Physical power thus reflects an ability to determine the behavior of the object. There are circumstances where such power is also exercised over a person, i.e. direct force changes the motion of an individual. For example, one can exercise power over another by pushing them in a certain direction. While this mechanism of power does occur, it is significantly different than the way power in social systems is often described. In social systems, it is typically assumed that each individual continues to be autonomously directed, but the control over an action that is performed can be transferred from the individual who is performing that action to another individual who wields control. This concept of the transfer of authority over decision making is central to the theoretical framework of power we describe here. 

When we consider the role of power in social collectives, power reflects a transfer of authority over decisions about actions by many individuals to one or a few individuals, or alternatively to a social abstraction such as laws or institutions. Such a transfer of authority gives rise to large scale social actions. The scale of those actions is to be considered the subject of a dynamic theory of the exercise of power by various authorities. Our framework provides a basis for such a dynamic theory.

\section{Formal framework for power}

In presenting a scientific analysis of a social construct such as power there is a need to provide formal justification for the model constructed as well as a presentation that can be understood at an intuitive level. These are linked to observations that begin from quite general ones and then proceed to progressively more detailed validation. This section provides a more abstract formal approach to arrive at the same framework for power as the subsequent sections and can be omitted at a first reading by those wishing for the more intuitive approach. 

We consider an individual as having the ability to make decisions about multiple possible actions, i.e. to choose over time a sequence of actions to perform of many possible such sequences. We consider the choices by the individual to be guided by ``motivations'' which are distinct from individual decision making about specific actions in their generality, i.e. motivations guide many decisions and their modification changes the overall pattern of decision making. The distinction between motivation and specific decision making is only in the abstraction of the model of decision making, i.e. a motivation is an aspect of the causal reason for many different decisions rather than a single one.

We define power as the transfer of decision making over one individual's actions from that individual to another individual (a leader). We are not primarily concerned about the mechanism of that transfer of power, i.e. the forms of communication that are present to facilitate or enable it. We will be particularly concerned with the case where the leader has power over multiple individuals rather than a single individual. 

In building our framework of power, we consider how motivations of an individual lead to power over the actions of that individual by another. Such a transfer of power can itself be considered an action and therefore should be understood from motivations of the individual. Human motivations are often limited in scientific discussions to striving for self-benefit (i.e. economic rationalism). We consider a more general concept of an objective oriented motivation---actions that are causally connected to valued conditions, i.e. values---with self-benefit as an important and common example of such values due to its self-consistency through selection for persistence through relevance for survival.  We allow for self-benefit or measures of achievement of values associated to an action to be measured both in a positive and negative sense, with terminology such as harm or cost for the negative direction.

Beginning from a model of an individual who performs actions for self-benefit, decision making is motivated by the individual's understanding of benefits and costs to the individual. This decision making can be influenced if the transfer of power to the leader has the ability to change the apparent payoffs of the set of actions that are possible for an individual to make.

We therefore consider four ways a transfer of power can be motivated. First, by the existence of an underlying expectation of improvement of benefit from the very nature of a transfer of power. Second and third, by the leader directly increasing the apparent cost or the apparent benefit to the individual of particular action options, leading to those actions being avoided or chosen due to those changes in payoffs. Fourth, by changing the evaluations of the payoffs by the individual, e.g. through changing values. These four options map onto the four different forms of power that we will consider (political, military, economic, and ideological). We note that the motivation for the fourth (power to change values) can arise self-consistently from within the values themselves, and has a potential to be inherently recursive, a key property of this mechanism for motivation.

\section{Political power and group collective activity}

In order to understand power we must also understand why group collective activities may serve purposes in social systems. In some contexts, there exist beneficial activities that can be performed by joint action of a group, which individual action cannot perform. Examples include successful hunting of large animals (i.e. attacks by multiple individuals may kill a large animal under circumstances in which an attack by one individual cannot), time bounded searching (i.e. search and rescue), and dominance in military conflict. More generally, it is widely believed that a large variety of problems can be solved by putting an individual ``in charge'' of the actions of multiple individuals. This belief is consistent with complex systems theory. When one individual is able to determine the behavior of multiple individuals comprising a collective, this enables the behavior of that collective to have a scale and complexity that may be needed for the successful performance of a task \cite{dcs,multiscalevariety}. We note that it is possible to show that the benefits of such individual authority are bounded under highly complex conditions where effective distributed control is more beneficial. However, there are also circumstances under which the expectation that putting an individual in charge will have benefits is justified. Under these circumstances it makes sense for an individual to desire to be part of a group that transfers authority over individual decision making to one individual, i.e. an individual chooses to be a member of a group rather than ``going it alone." For our purposes it does not matter if this results from evolutionarily derived (instinctive) behavior or a socially learned one (conscious or not). We call this type of transfer of authority for an implicit purpose of collective (and individual) benefits ``political power.'' Thus we see that it can be important to have a leader to perform effective collective action. 

While collective action may have inherent benefits, individual differences may give rise to different levels of decision effectiveness for political leaders. Thus, it may be important which leader is given that power. How a group determines which individual is given political power is a key aspect of how we understand group activities generally and systems of collective governance specifically.

We associate our understanding of political power with the conventional concept of political leaders. This is consistent with the observation that the mechanism of selection of such leaders can vary and does not have a specific origin that is characteristic of other forms of leadership. Other justifications for this association will arise from specific examples.

\section{Military and economic power} 

Two other kinds of power can be readily identified as the result of threat or reward, the proverbial stick and carrot incentives. In the former, a direct threat results in an individual choosing to transfer authority to avoid potential harm. In the latter, transfer of authority is done to achieve a direct benefit of a specific action. These are distinct from the general transfer of authority represented by political power in that they arise from specific individual assessment of sufficiently well defined negative or positive consequences of alternative options.

An institutionalized version of power that arises from threats involves a military or police force, so we term it ``military power.'' Actions that arise from associated benefits rather than threats are often described as voluntary and contrasted with threats as being coercive, and are considered as the subject of economics. We therefore use the term ``economic power.'' Our focus is not on all economic transactions, but rather on transactions that involve actions by one individual (labor) in response to benefits (wages). 

We note that the decision of an individual to accept a benefit depends on the relative benefit associated with alternatives that are available. Considering the relative benefit makes the distinction between voluntary and coercive concepts less clear than is often assumed in economics. I.e. if the only alternative to a transaction is negative the decision to transact is coercive in much the same way as a direct threat. For example, the alternative to wages may be starvation and therefore the choice made is essentially based upon a threat. Considering an individual who has control over a critical resource, i.e. water, such control can enable economic power that is coercive. Thus, the distinction between voluntary and coercive arises because of the nature of the alternatives. When there are, in some sense, adequate alternatives the exercise of economic power is different from coercion. Military power is by definition dependent on an adverse alternative; economic power is dependent on adverse alternatives only in so far as there are not better alternatives. In the study of economics it is assumed the alternative conditions that might be coercive are not the responsibility of the benefit provider. Accordingly, it is not the provider that is the coercive agent, which distinguishes economics from coercion. At the same time, the analysis of behavioral choices and actions of an individual must consider the set of options that are available to them and the causes of those options. 

\section{Values and ideological power}

It is apparent that underlying the exercise of economic power are the values of the individuals to whom benefits are offered. Values are embodied in the ``utility'' of something to an individual as an essential primitive concept in economic theory. Values are also essential to political power, as the acceptance of leadership, and more specifically the choice of a leader to whom authority is given, depends on the values of those who are voluntarily transferring authority to the individual who becomes the leader. Perhaps less obviously, values also underly the exercise of military power. While the threat of pain or death is an important motivation for accepting a transfer of authority, individuals may refuse to heed threats, may withstand pain, or even accept death, based upon values. The ability of different individuals to adopt different values and change them over time, and crucially to transfer authority over the setting of those values, leads to an additional form of power, ``ideological power.'' Ideological power in its purest form does not act directly to control actions but rather controls values that in turn control the decisions about actions. Ideological power interacts in important ways with other forms of power. For example, by changing values it can reinforce or undermine economic, military or political leaders. By this definition, ideological power can be seen to include both ideological value systems and the individuals who exercise authority over/within those systems. It is also apparent from the definition that ideological power encompasses religious value systems but also systems that are not generally included as religious, including patriotism and secular humanism. Value system leaders include all of those who provide narratives of meaning and significance about the world. In a secular context, this includes authors, reporters, entertainers, producers and teachers. There are embedded values in any fiction or non-fiction story and reporting, including what is important and how it is described. 

\section{Institutions and conventions with power}

Laws, institutions and other social conventions should be understood to serve two (equivalent) roles in our framework. First, they serve as the ``system of power'' within which power is exercised. Second, they can be understood to circumscribe the power individuals exercise at a particular time. In effect their presence assigns power to historical figures and processes which are abstractions of individual authority. The mechanism of authority of those abstractions must be embodied through individuals with current power whose authority can be understood in relation to those abstractions (e.g. specific individuals who have interpretive or enforcing roles) as well as the public. The values of the public in assigning authority (i.e. the authority they confer on abstractions and current leaders) constrains the degree to which leaders can exercise power in relation to those abstractions. In an important sense, laws, institutions and social conventions as values are incorporated in the religion/ideology of the populace. 

Thus, for example, religious (and other) leaders may be constrained in their actions by codified religious values. Political, military and economic authorities (as well as religious ones) may be constrained by legal systems, institutions and conventions based upon historical precedent. When leaders test the constraints of those codified systems, potentially violating or changing them, the response of other leaders or the populace determines whether they are able to do so.

\begin{figure}
\includegraphics[width=5cm]{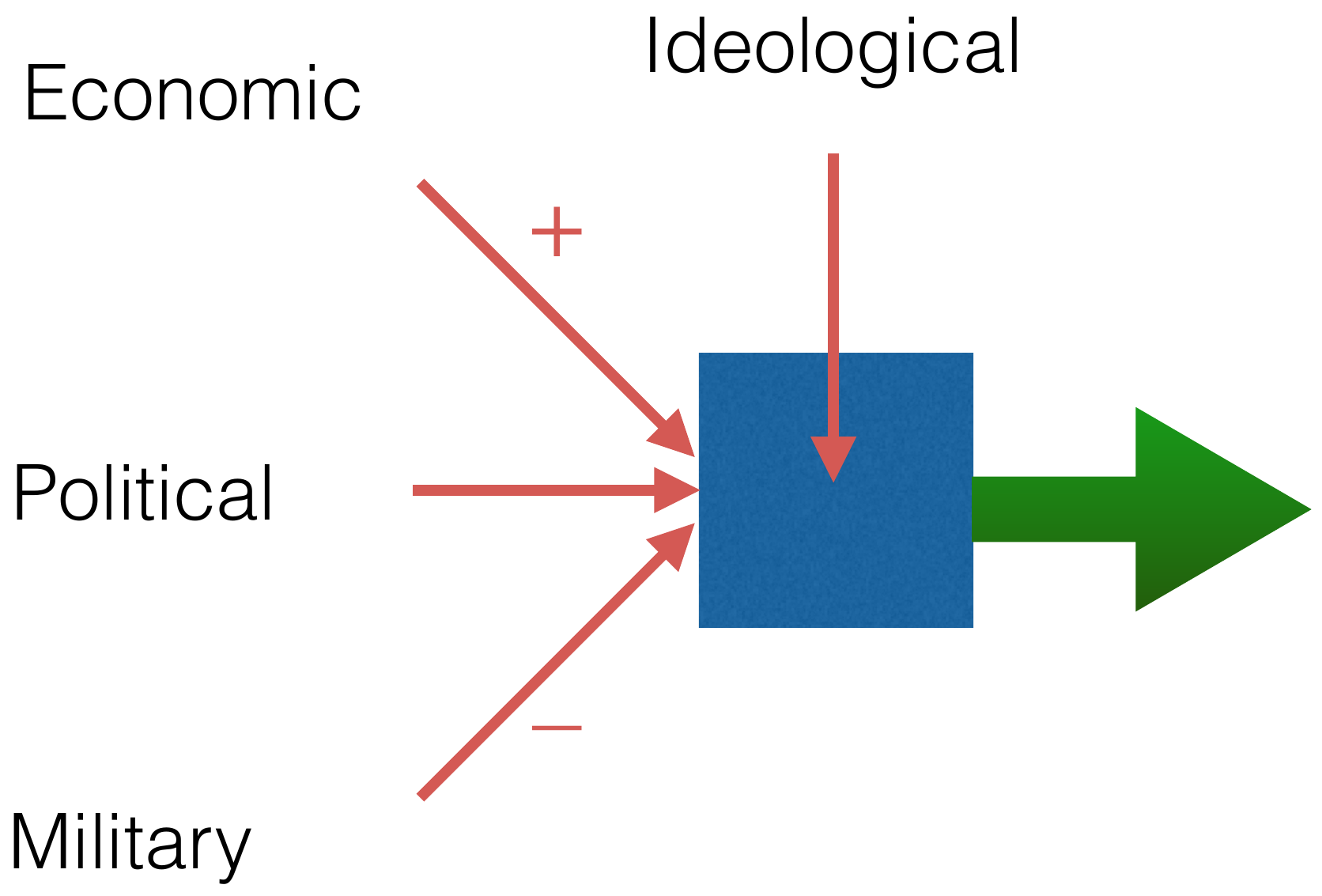}
\label{fig1}
\caption{We identify four types of power that enable one individual to control the actions of others. Economic and military power arise from rewards and threats, political power
from the expectation of group benefit from central decision making, and ideological power from transfer of authority over values.}
\end{figure}

\section{Graphical representation of an individual}

Fig. 1 provides a schematic diagram of an individual showing political, military, economic and ideological power as consistent with a simple model of individual behavior consisting of (1) a tendency to transfer authority to others (political power), (2) a desire for benefits (economic power),  (3) avoidance of harm (military power), and (4) a system of values, over which authority can also be transferred (ideological power). This elementary diagram of how the decision making of an individual is controlled and the actions that result can be used to illustrate various types of power systems. A power system characterizes the way aggregation of actions occur where different groups of individuals are controlled by competing powerful individuals, potentially with different types of power. 

\section{Graphical representations of aggregation (multiscale analysis)}

We can use the individual level representation to construct collective representations of social systems in which power is exercised. An example is shown in Fig. 2 in which a military leader, e.g. a dictator, uses military power to exercise control over the populace. The schematic not only shows that power is exercised over the populace but explicitly includes the role of individuals in the military to control the public, and the control of the leader over the military that gives rise to the control of the leader over the public. As is shown in the schematic, the relationship between the leader and the military may involve multiple channels of power, including values (ideological, i.e. loyalty, obedience), political (authority transferred for group benefit, i.e. for the military group or for the country as a group), military (i.e. coercive, threats that might also originate from other members of the military [not shown separately]), or economic (i.e. wages). Note also that economic power over the leader is exercised by the populace as a whole, i.e. through the flow of taxes to the leader. This power is not the same as that of an individual exercising power due to its distribution through the populace (but  suggests the importance of such reciprocal power, especially if wealth becomes concentrated, or in the event of collective action). The complexity of the diagram illustrating a real world system that involves military power is consistent with a need for embodying the power in relationships between individuals used to create that power. Thus, saying a military leader exercises power is not enough, the military force through which the power is being exercised is part of the representation of that power. A variety of other kinds of power structures can be similarly illustrated by such diagrams (e.g. Fig. 3). 

\begin{figure}
\includegraphics[width=7cm]{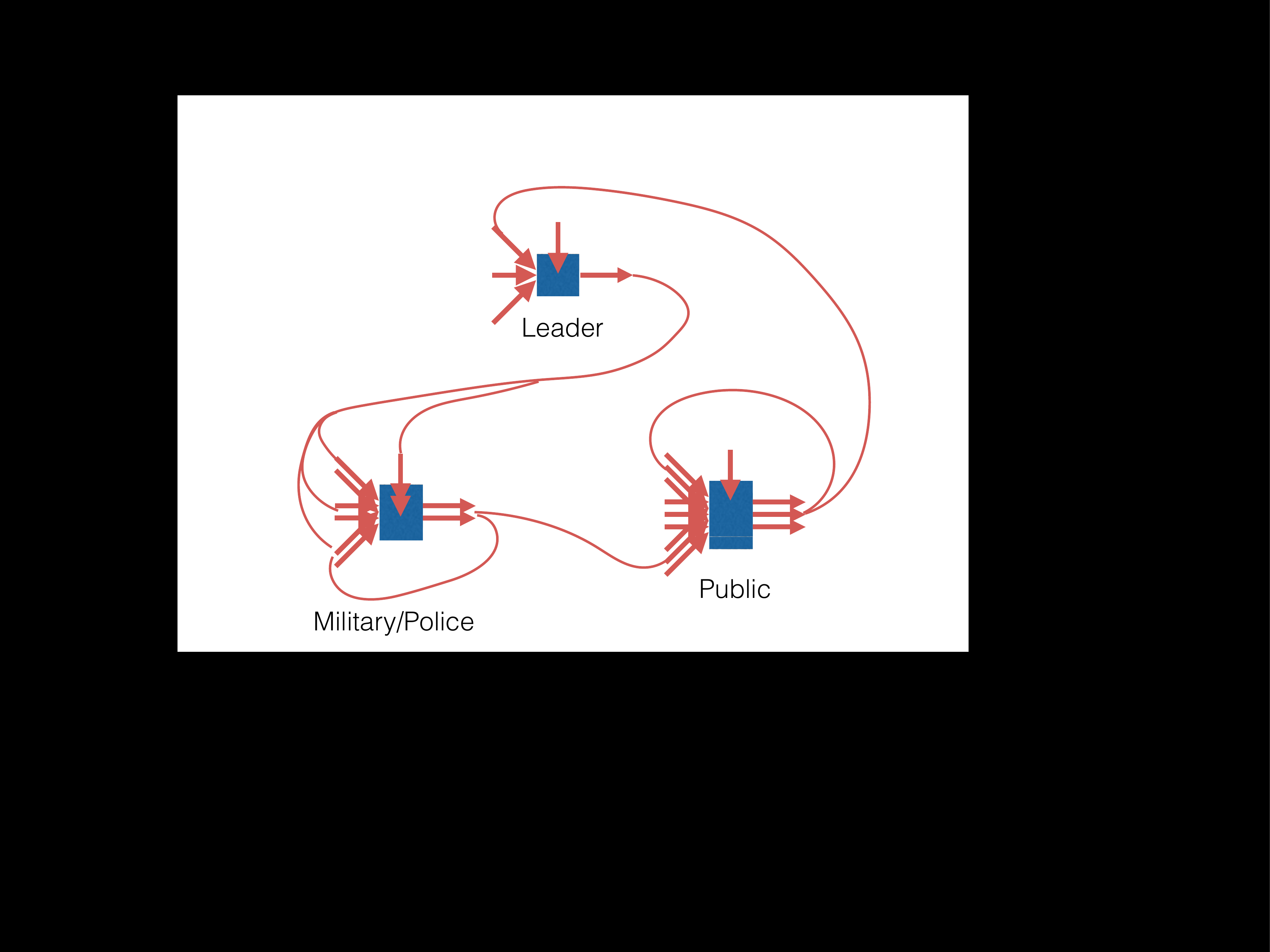}
\label{fig2}
\caption{Schematic of a social system in which military power dominates. The leader uses its control over a military or police force to exercise power over the populace, extracting resources from the populace that it uses to reward and thus control the military force.}
\end{figure}

\begin{figure}
\includegraphics[width=7cm]{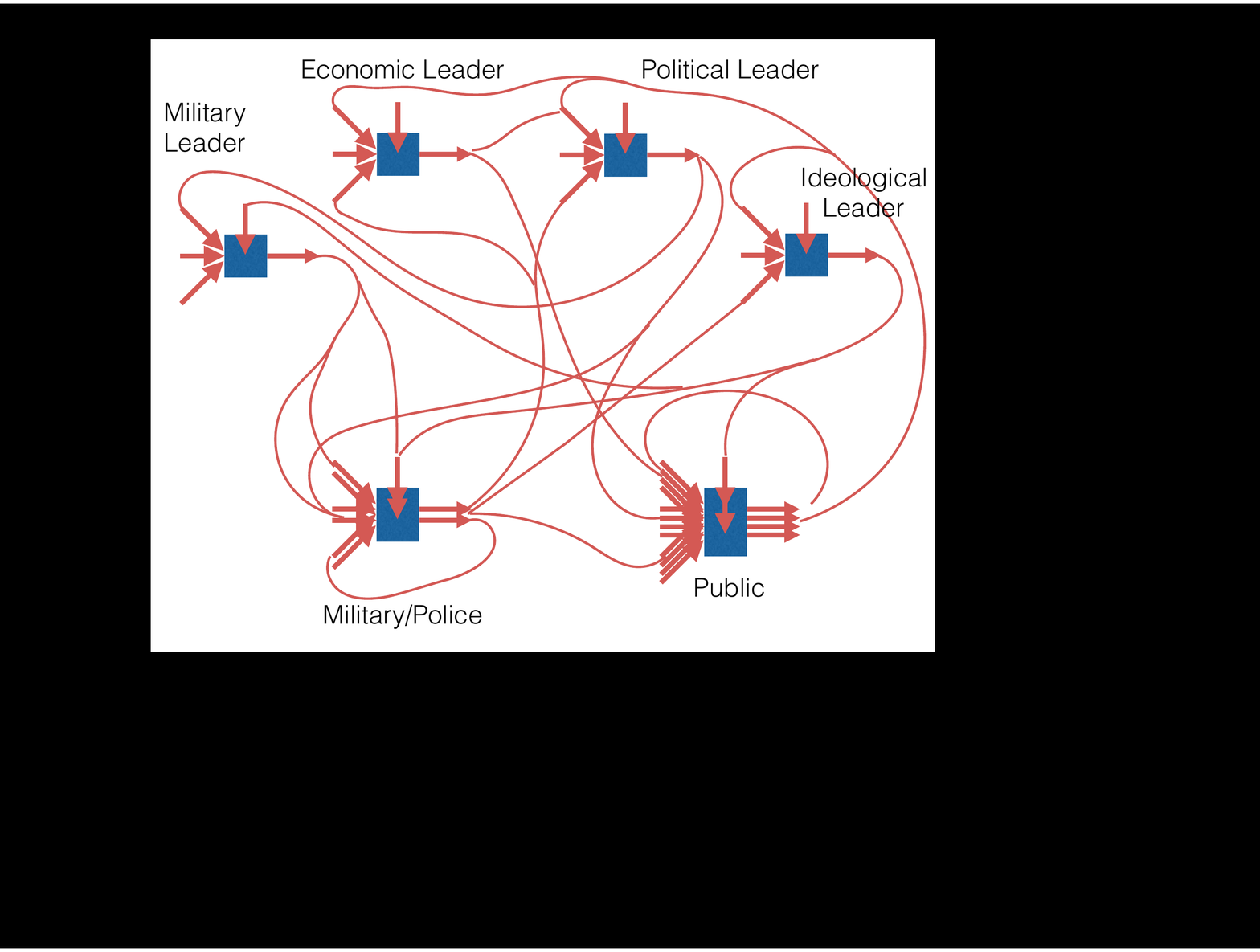}
\label{fig3}
\caption{Schematic of a social system in which leaders with each of the four powers play a role. Many but not all of the roles that power can play are shown. Military forces are shown in addition to the military leader. However, in a more complete figure there would also be various political operatives, corporate managers and clergy and lay leaders.}
\end{figure}

\section{Notation for representation of social power systems}

In this framework, a society is described by a set of power relationships that comprise a kind of network. We can define a society as such a network that does not decompose into independent subgroups. A quantitative representation of a society is a list of individuals or groups with a specified number of members and the associated individuals or groups that exercise power over them:
\begin{equation} \begin{array}{ll}
(n_1, & \text{name}_1, \text{name}_{1,p}, \text{name}_{1,m}, \text{name}_{1,e}, \text{name}_{1,r}) \\
(n_2, & \text{name}_2, \text{name}_{2,p}, \text{name}_{2,m}, \text{name}_{2,e}, \text{name}_{2,r}) \\
... \\
(n_k, & \text{name}_k, \text{name}_{k,p}, \text{name}_{k,m}, \text{name}_{k,e}, \text{name}_{k,r}) \\
\end{array} \end{equation}

\noindent where $\text{name}_i$ is an individual's name or a group name with number of members $n_i$. Each of the other names in a list is the identifier of an individual or group exercising power over that individual, i.e. a leader of a particular type, in the order of political (p), military (m), economic (e), or ideological (r). A null or self-reference of a leader represents the case where that type of leader is not exercising power over the individual, or does not exist. The properties of the social system depend on the numbers of members of groups and the structure of the network of power links. For analysis of the structure the labeling by names is not essential and may be replaced by ordinal numbers that index that individual or group. 

The formalism can be extended to include laws and other forms of constraints on leaders, i.e. constitutions, social conventions, religious documents, or other codified guides. The simplest way to include them is to name them and to allow the number of individuals associated with that element of the system to be zero. This is similar to the inclusion of extrinsic forces on the system in general models of influence networks \cite{chinellato2015dynamical,harmon2015anticipating}

For example, a single leader with a military based dictatorship with 1,000 military members and simultaneously religious authority over a million citizens might be represented as:
\begin{equation} \begin{array}{ll}
(1, & 1, 1, 2, 3, 1) \\
(1000, & 2, 1, 2, 1, 1) \\
(1000000, & 3, 1, 2, 1, 1) 
\end{array} \end{equation}
Based upon historical information about the role of Roman Caesars in political, military, and economic leadership, this might be considered a simplified model of the Pax Romana system (see below). 

A dictatorship with a separate religious leadership with 100 clergy might be represented as:
\begin{equation} \begin{array}{ll}
(1, & 1, 1, 3, 4, 2) \\
(100, & 2, 1, 3, 1, 2) \\
(1000, & 3, 1, 3, 1, 2) \\
(1000000, & 4, 1, 3, 1, 2) 
\end{array} \end{equation}

The notation can be extended to represent systems in which more than one individual or group exercises partial power over a particular individual or group, by making the substitution $\text{name}_{k,s} \rightarrow \{\text{name}_{k,s,i},w_{k,s,i}\}$ with $s$ in the set $\{p,e,m,r\}$, $i$ an index, and $w_i$ the weight of the respective influencer with partial power. This would then be a network representation of how individuals are influenced by multiple other individuals or groups. The network representation we are identifying has additional structure compared to a standard influence network \cite{axelrod1997dissemination,chinellato2015dynamical,harmon2015anticipating} by virtue of distinguishing the different categories of influence, and allowing nodes to represent aggregated groups rather than individuals. The representation is different from that generally considered in network theory of generalized mutual influence with a specified structure of the influence network. It has more distinct structural elements, but collapses the number of individuals of the society and the relationships among them onto a concise and simpler form. This simplification, in which individual agents have power over groups, may be useful in analysis. 

The framework developed thus far in this paper describes a system of roles rather than of individuals within those roles. Missing is the mechanism by which individuals are placed into those roles. Those mechanisms are often integral to the power system itself. Indeed, in traditional political thinking the definition of a political structure is often based upon that mechanism. Thus a dictatorship or a representative democracy might in principle (though not usually in practice) differ only in the way the individual in power gains that power rather than the powers that that individual exercises.  We can generalize the framework we have discussed to include a characterization of the processes by which individuals attain power. There is, however, enough to discuss in the context of the power structure itself for a first article so we defer much of the discussion of the processes by which power is attained to a follow-on 
article, only mentioning it when essential. Instead we focus on the dynamics of power systems, the way that the structure of power in society, as described by the representation we are using, changes over time.

The dynamics of power systems can be considered in terms of the representation by identifying (1) illustrative examples of power systems and their historical trends and 
(2) dynamics of power systems through (a) competition between systems, (b) changes of the structure of a system by changes in the types of roles present (i.e. whether a single individual plays multiple power roles), or the power relations among powerful individuals, and (c) changes in the values of individual that plays a certain role. Constructing examples requires the ability to refer to history, a process fraught with scientific issues but one which cannot be avoided in this context. 

\section{Use of historical consensus narratives as scientific data}

Collecting data about social systems for scientific analysis presents many challenges. Among these is the implementation of controls on the objectivity of observational data. In order to discuss our models of power it is helpful to have information to compare with across historical periods. Data sources are often limited to observations by individuals that are then transmitted through a chain of communications to the present. Establishing historical data ``facts" for scientific analysis is therefore  problematic. However, where information affects many individuals, the historical communication of information is more reliable than the communication of information that pertains to individual actions that are less influential. This reliability arises because of the redundancy of the information, which promotes valid transmission through unreliable communication channels \cite{shannon}. In this case, the communication channel is the historical transmission of social information. We therefore consider consensus historical narratives that describe events or individuals as to their influence on many people as having an appropriate, if limited, level of validity for inclusion in scientific analysis. It should be apparent that the validity of even such events may be challenged. The opportunity for such challenge exists and notwithstanding the possibility that individual events may be found to be incorrectly reported, a scientific theory may be compared to existing historical narratives. We will proceed to consider the way that historical power systems can be interpreted within the proposed power framework. Mann's four volume treatise  \cite{michael1986sources,mann1993sources,mann2012sources3,michael2012sources4} can be consulted for a much more extensive discussion of the interplay of the four types of power around the world and through history. 

\section{The exercise of power by leaders}

A power system, by definition, means that leaders have the ability to control what other individuals in the system do. The actions of individuals that are controlled can serve many different purposes. Actions that can be controlled can be considered a ``resource" like land or raw materials. Leaders will control the actions of others to achieve what they themselves consider benefits. Thus the value system of the leaders is an essential attribute of leadership and its role in society. The premise of individual motivations implies a central question is whether the values of the leader result in self-serving benefits or benefits to those who are controlled by them, or other groups. The way leaders are selected plays an important role in the leader values that are likely to be present. Achieving benefits that are desired by a leader, including self-serving benefits, is a motivation for individuals to become leaders. Given multiple leaders, the ability to control is limited by or shared with other leaders. This limits the ability of one leader to achieve what they believe is of benefit. Therefore, one of the likely goals of a leader is to increase their power, so as to increase their ability to serve what they consider to be a benefit. As discussed later, this is a generic part of the dynamics of power systems whether or not the leader is self-serving. 

\section{Illustrative historical examples of power systems}

The four different types of power can interact with each other in different ways in a specific system. Whether types of power are combined together within an individual or distributed among distinct individuals and the way powerful individuals exercise power over each other, so that one type of power might dominate others, distinguishes different systems. 

The existence of different forms of power must be combined with social separation of these powers in order for them to be recognizably distinct at an aggregated level. Historically, they are not, however, completely separated but in different systems they become segregated in various ways. Understanding the structural forms of power distribution and their dynamics is important scientifically and relevant historically. Perhaps a good way to think about the way this works is like identifying primary colors in a picture. The elementary nature of these forms of power allows for concentrated patches but does not preclude various forms of mixing that are considered by Mann as networks of power. 

Here we limit our discussion to a few observations and general statements of relevance to historical dynamics of power systems. One observation that appears supported by the historical record is that power has become disaggregated over time in large countries compared to ancient Rome and other early empires. A common origin of the growth of ancient empires was through military conquest. This naturally led to military leaders that dominate other forms of power. By contrast, in modern military dictatorships,  ideological power (often religious) and to varying degrees economic power are often separated as they were not in ancient ones.

In the ancient Roman empire during the Pax Romana period, a Caesar exercised all forms of power, i.e. political, military, economic and ideological leadership. In particular, Caesar was considered a god during this period. At the beginning of this period, the classical Roman civilization went from a republic dominated by military action to a society with essentially three groups \cite{michael1986sources}: the emperor, soldiers, and the mass of the people---including slaves (an estimated 20-40\% of the population at the beginning of this period) and nominally free peasants that merged together into serfdom. A small number of people, the leadership of the military, might also be considered a distinct group. The budget (combining that of the state and that of the emperor, which were not substantially distinct) was 70\% for military and 15\% for the dole to the Roman public, presumably the minimum that could be given and maintain order. In contrast, US military spending is 20\% of total federal spending / 12\% of federal, state and local spending. The Pax Romana period was as pure a version of a coercive society as may be possible with human beings. Its success in space-time extent can be understood from the stability of such a system given the value system of the time. 

The advent of the major religions led to partial, though surely not complete, separation of ideological power from other forms of power in many areas of the world. Later secular ideologies, including communism, secular humanism, and more generally patriotism, when they arose became more closely associated with political power. Theocracies embody the possibility that ideological power becomes dominant, exercising also all other forms of power to a greater or lesser degree. This is apparent in the periods of Christian Inquisition and the Crusades, the Islamic conquest and caliphates, and the modern state of Iran as key examples. The dominance of religious leaders over other forms of power can be distinct from military leaders exercising religious authority due to the origins and primary mechanisms of their power. Communism suppressed competing ideologies including religions. Modern governance structures that appear to be either absent a religion or having adopted pluralistic separation of church and state may be better characterized as variations on patriotic secular humanism. In this context, the separation of church and state enables sufficiently compatible religious systems to coexist with overlapping roles in determining values. 

Confucianism provides an important example for our purposes. Mann's characterization\cite{michael1986sources} suggests that ``Confucianism's role is thus largely one of morale boosting: it introduces no principles of ideological transcendence." Our analysis suggests a different perspective focused on the role of the value system itself. Confucianism has a transparent and self-declared purpose of shaping the values of leaders away from self-interest and toward benefit to others, as well as loyalty of subjects. According to its principles, having virtuous rulers is central to stable governance. In this it directly targets the essential role of leadership values: When leaders adopt values that serve the public, they are supported by the public. Leaders would adopt it because of the public loyalty it inherently as well as explicitly promotes. When the public considers leaders as not fulfilling those values, leaders would be replaced. Indeed, leadership changes did occur and Confucianism dominated Chinese values for 2,000 years until the advent of Communism. Confucianism also embodied an extreme version of a codified value system, where written abstract ideas are the essential version of the ideological power. The common knowledge of these ideas and their transmission through educational institutions is, to first order, the extent of institutionalized power structure. 

Since the industrial revolution and through the 20th century, the relationship between economic and political power was the subject of much of the global ideological conflict, with different degrees of dominance of political leadership over economic resources and thus economic power as reflected in capitalism, socialism, and communism. These ideologies are in part descriptions of power systems, though the relationship between the ideological values and the reality of the power system effects may not be direct or complete because of limited understanding of the real world consequences of their implementation. That people value and expect certain outcomes of power systems does not make them true properties of those systems.  

In an idealized well functioning modern democracy, the four domains of power are considered to be separated. Each of the powers dominates in its own sector of public activity, political leadership making policy, military and police leadership ensuring security, economic leadership making production and investment decisions, and religious leadership promoting altruistic and patriotic values. Political power exercises limited control over economic power through regulation, and over military power through control over resources, 
and also through ultimate command and the declaration of war. 

While separation of the domains of power is considered the ideal, the exercise of power of economic leaders over political power through regulatory capture is widely considered to be found in capitalist systems that emphasize independence of economic activity from political control, including in the US.  Corruption is widely considered to be present in the places where political power has greater control. We note that US policy has long opposed corruption. This may seem to be paradoxical as corruption is superficially similar to the kind of influence that economic interests have over the US government. However, our analysis shows that there is a difference. Influence buying or regulatory capture occurs when economic power is being exercised over political power. In contrast, corruption reflects dominance of political power over economic power. Hence, economic leaders do not like corruption. The difference is subtle but real. The practical reason they are different is that regulatory capture enables ``special interests'' to exercise control over policy, while corruption enables a political power to extract money for a specific favor that need not provide systemic benefits, i.e. it is part of business activity (a kind of tax) rather than part of business policy. When political power is dominant (China today), political leaders are the primary beneficiaries of economic activity (i.e. wealth). When economic power dominates,  economic leaders are the primary economic beneficiaries and they have effective control over economic policy (US  today).

When one power dominates, that power gains authority over other forms of power, determining public decision making (political power), coercively promoting their control (military power), extracting wealth from and controlling economic activity (economic power), and promoting values that favor themselves (religious power). Under such conditions the boundaries between different kinds of power become blurred. 

Moreover, when one type of power becomes stronger than other types of power, individuals whose 
leadership role originates from one source of power become leaders with other types of power. Thus, when economic power dominates, wealthy individuals become political leaders. When political power dominates, political leaders become leaders of economic institutions, i.e. corporations. These two directions are apparent in the examples of the US and China today \cite{chinacorruption,UScapture}. Consequences are similar for military and ideological/religious leaders when they are dominant. 

It is important to emphasize that the populace exercises power in every system and this power is manifest in diagrams (Figs. 3 and 4) that show the populace exerting power over leadership. Collectively, the power of the populace is strong, generally stronger than any leadership, but individuals or even groups of the populace have respectively less power if they do not act in a unified way.

\section{Dominance of types of power}

More systematically, we can list the effects of dominance of one kind of power over another:
\begin{itemize}
\renewcommand{\labelitemi}{$\bigtriangleup$}
\renewcommand{\labelitemii}{$\bigtriangledown$}
\item Where military power dominates over
	\begin{itemize}
	\item political power: we have dictatorship, police state, loss of rule of law
	\item economic power: we have coercive exploitation, extortion (e.g. organized crime)
	\item religious power: we have generals as icons or gods (e.g. Rome)
	\end{itemize}
\item Where economic power dominates over 
	\begin{itemize}
	\item political power: we have regulatory capture, exploitation of natural resources and exploitation of workers (e.g. low wages)
	\item military power: we have coercive employment including prison labor and slavery
	\item religious power: we have aspiration for wealth and/or acceptance of subservience,
		 work is done for hoped for (prospect of) money and not for money
	\end{itemize}
\item Where political power dominates over
	\begin{itemize}
	\item military power: we have wars of conquest
	\item economic power: we have corruption
	\item religious power: we have patriotism, suppression of other religions
	\end{itemize}
\item Where religious power dominates over
	\begin{itemize}
	\item military power: we have persecution, inquisitions and religious wars
	\item economic power: we have cult behavior
	\item political power: we have religious law, prohibition, witch hunts (e.g. McCarthyism)
	\end{itemize}
\end{itemize}

Modern examples of relative but not complete dominance of each of the four forms of power can be found in Egypt for military power, US for economic power, China for political power, and Iran for religious power.

\section{Dynamics of power}

There are three kinds of dynamics of power: competition between power systems and their specific instantiations, competition of leaders for power including aggregation and inter-leadership relationships, and an individual leader's transformation of values that affects how they allocate the resources they have control over. 

While there are distinct aspects of these three forms of dynamics, there are also ways in which they all reflect the competition between leaders. Leaders compete for power within a power system, as well as to gain control over others that are currently part of a different power structure. The latter can be reflected in conflict between two systems with the same type of power, or two systems with different types of power systems. Individual leaders change their values and associated behaviors in ways that depend on how vulnerable or secure their control is 
in relation to other leaders. Still, dynamics may not be entirely reflective of leader competition for power, as, for example, power systems can compete even when they are not 
subjects of the same individual leaders.

\subsection{Dynamics of power systems}

One of the central aspects of the historical narrative is the changes that occur in power systems as reflected in the prevalence of system types. While our analysis of power is not exactly the same as the traditional categories of political systems, changes in political systems are changes in power systems. Thus the historical changes of prevalence of monarchies, dictatorships, democracy and communism reflect a competition between power systems. This competition occurs through gradual change within a country, conquest by one country of another, as well as by revolution and mimicry of power systems by one country of others. 

The outcomes of competition between power systems reflect the advantages that one power system has over another. Intuitively a historical progression among power systems arises because they provide increasing competitive advantages, a kind of evolutionary dynamic. Competitive advantage may arise from two different sources, which are not 
independent: advantage to the populace, hence their selection by the populace, and advantages in competition at the collective scale, in one or more of the categories of military, economic or religious conflict, as reflected in the mechanisms by which such conflicts are won.

In direct conflict between power systems where leaders of the same type and their followers compete, the winner is determined by qualities that can be readily identified, though other aspects of the system may play a significant role. For example, conflict between militaries is generally considered to be won by the scale of military forces as measured by manpower and firepower. However, factors such as military technology and production capacity may play a key role and unconventional warfare points to other factors of importance in particular contexts. Conflict between religions is won by proselytization and stubbornness (as well as ethnic conflict / massacres and conquest of territories). Conflict between economic systems is won by resources and their effective allocation (as well as innovation). Conflict between political power systems is won by popularity contest (as well as political maneuvering).

When we consider the relative advantages of different types of power systems, we should recognize that evolutionary advantages are context and tradeoff dependent and therefore involve subtle considerations and careful analysis. This is manifest in biological evolution through the diversity of biological organisms that are present in the world today. For example, it is commonly stated that a faster running cheetah has evolutionary advantage over a slower one, hence the high speed they achieve. However, not all animals are as or even at all rapidly running, nor are some able to run at all, and some can fly faster than a cheetah can run. This points to the context specific nature of selective forces. We may infer that such context specificity is also present in the evolution of social systems and their power structures. As a first step, however, we can identify some advantages of power systems in relation to each other by straightforward considerations. These advantages are to be understood as subject to clarification of the conditions under which they apply. For example: 
\begin{itemize}
\renewcommand{\labelitemi}{$\triangleright$}
	\item Military dictatorships provide order and are thus preferred over disorder, i.e. anarchy, as an absence of a power system. Dictatorships will be selected for over anarchy under conditions where large scale resources are available, including those obtained by conquest, and other external forces are not disruptive of the power system. This is similar to multicellular organisms, which have advantages over single celled organisms in the presence of large scale resources (e.g. food) that can be consumed, one of which may be mutual consumption, and because of the homeostatic environments that are formed within them.
	\item Monarchies are advantageous over dictatorships due to reduced conflict over leadership transitions. This will be selected for under conditions of multigenerational persistence of the power system. 
	\item Democracies and Communism as political systems are advantageous over monarchies due to the advantages associated with selection of leadership values and capabilities. They will be selected for under conditions in which leadership decisions must respond to an elevated level of internal complexity of the society (note however the limitations of hierarchical representative systems \cite{multiscalevariety}). 
	\item Free market based Democracies are advantageous over systems that exercise price and production control (e.g. Dictatorships and Communist systems) due to the advantages of economic coordination for complex economic activities. 
	\item In general, we can argue that competition among countries due to system capabilities and effectiveness and for popular support tends to aggregate power where large scale coordinated actions are needed, and disaggregate power through specialization of leadership and increased complexity of actions in the face of increasing complexity in actions. The latter is similar to the evolutionary development of cognitive capacities in animals through neural systems as being an evolutionary advantage in the context of environmental complexity of decision making.
\end{itemize}

Competition between groups involved in direct confrontation may reflect competition between power systems, when the power systems are different. It may also reflect competition between different instances of the same power system. In the latter case it may in part reflect competition between leaders for control over larger population groups, a topic that is further discussed in the next section.

Finally, we note that power systems may change not because of direct competition but rather because of changes in the conditions. Thus, discovery of natural resources and technological changes can change the resources that are present in the system. The changes in resources then affect power structures. For example, the industrial revolution in changing the means and organizations associated with economic wealth also created new power structures. 

\subsection{Dynamics of leadership}

Evolutionary dynamics are only one aspect of the dynamics of power systems, just as evolutionary dynamics are only one aspect of the dynamics of biological organisms. For clarity, among the others in biology are development, aging, damage, recovery, metabolic, cellular and molecular processes. 

Key to the dynamics of power systems are internal country competition and cooperation between leaders. Leadership competition tends toward consolidation of power as leaders seek increasing control as a means to achieve their objectives. Note that this internal dynamic of power aggregation is generally counter to the evolutionary dynamic that seeks to distribute that control for effectiveness of the system. Aggregation of power is limited by death, and thus mechanisms of inheritance are a key aspect of how power aggregates over time. 

In the process of power consolidation by leaders, different types of leaders can employ different methods. We can identify these methods from their type of power. In each case there is a mechanism of direct power use to take control, and a mechanism of cooperation that can result in power sharing, mutual strengthening and dependency. Specifically:
\begin{itemize}
\renewcommand{\labelitemi}{$\triangleright$}
\item Political leaders gain control through popular support for subverting or conditionally legitimizing other leaders. 
\item Military leaders gain control by threatening or by providing security for other leaders. 
\item Economic leaders gain control by buying out or providing resources for other leaders. 
\item Religious leaders gain control by proselytization of other leaders or by advocating on their behalf to others. 
\end{itemize}
When a particular type of leader gains control they not only change the balance of power between leaders, they change the nature of the power system by strengthening that type of power. Note that in all cases the differences between different types of methods used may be driven by values as well as circumstances.

Thus, for example, where leaders cooperate, economic power supports military power through taxation as part of social values promoted by religious leaders reflected in laws established by political power. Military power supports political power through preventing anti-leadership or revolutionary activity, as well as defending against foreign occupation and other threats. Ideological leaders promote values of leaders promoting economic, security and other benefits for the public, as well as values of the public (and leaders) that reinforce support of leaders. Political leaders implement laws and allocate resources for the public, as well as for economic growth and military defense, reflecting values promoted by religious leaders. 

There is an important distinction between leaders that cooperate with each other for mutual support and leaders that cooperate in support of the system of power itself. Military leaders might promote particular economic or political leaders, but this would undermine the structure of an economic system in which economic competition is the mechanism of promotion of economic leaders, and a political system in which political leaders are selected by public opinion and thus motivated and having capabilities in the promotion of public good. Similarly, economic leaders might promote individual political leaders by financial support, but this undermines the structure of the power system in a context in which political leaders are selected because of motivation and benefits that they provide to the public. 

As another example, where military power competes with economic power, military power can threaten economic power to obtain resources, or economic power can buy out military power, i.e. gain authority over it by providing support. Either way money flows to military power, and coercive control to economic power. If the extent of such competitive interaction increases, military forces may gain extensive control over wealth, or economic forces may gain extensive control over military actions to control workers by coercion instead of paying for labor. In this context we can see how values (ideology/religion) are an important aspect of the competition for power. For example, military values may limit the extent to which military leaders seek wealth. On the other hand, if it favors wealth, then military power may threaten economic power to obtain dictatorship control over economic resources. The competition also occurs in the context of popular support. For example, military power and economic power compete if they are both trying to gain political power.

We can see that the different kinds of power cooperate or compete in different structures. In a cooperative context:
\begin{itemize}
\renewcommand{\labelitemi}{$\triangleright$}
\renewcommand{\labelitemii}{$\circ$}
	\item Military power provides for security and enforcement of laws, i.e. social order, which promotes the framework in which 
		economic, political and religious power is exercised, 
	\begin{itemize}
		\item it enforces rules of ownership and protects the wealthy from crime, 
		\item it provides security for the political leaders against opponents,
		\item it provides security for religious leaders,
		\item it provides security for the public. 
		\end{itemize}
	\item Political power provides 
	\begin{itemize}
		\item laws that protect military, economic and religious leaders,
		\item laws that enable economic leaders to make economic decisions and exploit resources, 
		\item subsidies for economic activity, for the military, and for the populace.
		\end{itemize}
	\item Religious power provides values that promote
	\begin{itemize}
		\item public respect for economic, political and religious leaders and their decisions,
		\item public actions in support of collective causes,
		\item distribution of resources for the well being of the society and the public in particular.
		\end{itemize}
	\item Economic power provides management of resources and labor and coordination of processes 
	\begin{itemize}
		\item for the public good, 
		\item in support of military actions and religious communications,
		\item in support of political figures reflecting successful outcomes of policy decisions that promote economic growth and social benefits. 
		\end{itemize}
	\end{itemize}
In each case we can replace positive with negative interactions for systems with antagonistic interactions. Antagonistic interactions undermine leadership. Undermining is distinct from, but may be part of, interactions that lead to takeover by one power of another.

\subsection{Values of leaders and its dynamics}

Power can serve for the benefit of the group, including the individuals that are subject to power. It is also apparent that power can serve for the benefit of the individuals who are wielding power. Which is the case?

Decisions by an individual in power may lead to varying degrees of benefit for the collective and for the individual in power. Values of a leader determine whether their decisions are intended for the benefit of themselves or for the collective that empowers them (or another group). Thus the benefits of power can either serve an individual or the collective. While intentions and outcomes are not always aligned, under conditions in which consequences of actions are clear, intent corresponds to outcomes: Capability and values combine to determine societal and individual benefits. 

Quite generally, the values of an individual affect (1) what he does for himself and 
(2) what he does for others (others may be distinguished in group categories). For an individual who wields power and controls actions of others and receives resources from others the values of that individual affect (1) what he uses/takes (allocates) for himself and 
(2) what he uses/gives (allocates) to others (in group categories). This includes both immediate use and allocation for future use.

Thus, all types of leaders may and often do use power for self-benefits, including self-benefits that are considered to be counter to the idealized values of the power system. 
Self-benefits may be considered through biological and social needs and desires. This is manifest in financial gain and associated luxuries, and sexual misconduct as reflected in ``scandals.'' Note that this includes religious leaders even when the values they profess and espouse for others contradict such actions.

We can identify a theory of individual transformation of values in the expression ``power corrupts.'' This statement embodies a dynamic concept. Specifically, that there is a change in individual values as a result of the wielding of power. The change shifts the behavior of the individual toward serving him or herself rather than others. Implicit in this dynamic is the existence of alternatives, i.e. that an individual wielding power may not be serving him or herself. We see that this simple expression captures a remarkably rich set of concepts. It suggests that persistence of power leads progressively to more self-allocation and use of power for self-benefit.

\section{Conclusions}

We have identified four types of power based upon the underlying structure of human motivation and characterized the historical nature of power systems and the dynamics of power in terms of them. Societal benefits from establishing leadership for collective action can be identified. However, the essential paradox of transfer of power to a leader is that the purpose of the transfer of power as far as a group is concerned is for benefit to the members of the group, but the opportunity is present for the leader to exercise power for 
self-benefit. Power dynamical processes include (1) the dynamics of change of the relative benefit assignment to leader and group, (2) the competition and cooperation among different leaders, including different types of leaders, and (3) competition among power systems. It can be expected that the dynamics of individual leader values favors increasing self-benefit over time (power corrupts), the dynamics of power shifting may under some conditions favor group benefit but also power aggregation, and the evolutionary dynamics of groups and power systems at least under some conditions favors distributing power and group benefits, though group benefit may not equally serve every group member. Historical processes appear to result in a shift toward distributed power systems, though individual countries today continue to be dominated by specific types of power including political (e.g. China), economic (e.g. US), military (e.g. Egypt) and religious (e.g. Iran) leaders.

\end{document}